  \documentclass[12pt,epsf,aps]{revtex4}
\usepackage[dvips]{graphicx}
\usepackage{latexsym,amsmath,amssymb,bm,euscript,eufrak}

\begin{document}

\title{A Proposal to Use Neutron Scattering to Measure Scalar Spin Chirality Fluctuations in Kagome Lattices}
\author{Patrick A. Lee}
\affiliation{Department of Physics, MIT, Cambridge, MA 02139, USA}
\author{Naoto Nagaosa}
\affiliation{Department of Applied Physics, The University of Tokyo, 7-3-1, Hongo, Bunkyo-ku, 
Tokyo 113-8656, Japan \ \
and Cross-Correlated Materials Research Group (CMRG), 
and Correlated Electron Research Group (CERG), RIKEN-ASI, Wako, Saitama 351-0198, Japan}
\date{today}
\begin{abstract}
In the theory of quantum spin liquids, 
gauge fluctuations are emergent excitations at low energy. 
The gauge magnetic field is proportional to the scalar spin 
chirality $\bm{S}_1 \cdot \bm{S}_2 \times \bm{S}_3$. 
It is therefore highly desirable to measure the fluctuation spectrum of the 
scalar spin chirality. We show that in the Kagome lattice with a 
Dzyaloshinskii-Moriya term, the fluctuation in $S_z$ which is readily 
measured by neutron scattering contains a piece which is proportional to the chirality fluctuation.
\end{abstract}

\maketitle

It has long been suspected that the spin 1/2 antiferromagnetic Heisenberg model 
on the Kagome lattice may support a spin liquid ground state, i.e., a singlet ground 
state which has no Neel order due to quantum fluctuations.\cite{1,2}  
Several years ago the compound ZnCu$_3$(OH)$_6$Cl$_2$ (called Herbertsmithite) 
where the Cu$^{2+}$ ions form $S = 1/2$ local moments on a Kagome lattice was 
synthesized.\cite{3,4} Despite an exchange constant $J$ estimated to be $\sim 200$~K, 
no magnetic order was detected down to 30~mK.  Recent neutron scattering shows that 
the spin excitations are gapless and form a broad continuum.\cite{5}  Thus Herbertsmithite 
has emerged as a strong candidate for the spin liquid state.  However, for reasons described 
below, much remains unknown about this material and the connection with theory is 
tenuous at best.  There is thus a strong need for more experimental probes to help establish 
the nature of this state of matter.

Theoretically it has been proposed by Ran {\em et al}.\cite{6} based on projected fermionic 
wavefunctions that the ground state  is a U(1) spin liquid, with spinons which exhibit a gapless 
Dirac spectrum.  On the other hand, recent DMRG calculations on finite size cylinders show 
strong evidence that the ground state is a Z$_2$ spin liquid, with a substantial triplet gap.\cite{7}  
However, the nearest-neighbor Heisenberg model appears to be a very delicate point, 
because a small ferromagnetic next-nearest neighbor exchange $J_2 \approx -0.01 J$ is 
sufficient to destabilize the Z$_2$ state.\cite{8}  Meanwhile, more detailed projected 
wavefunction calculations show that the Dirac state is surprisingly stable.  Furthermore, 
the application of a couple of Lanczos steps produces an energy quite competitive with 
the energy of the Z$_2$ state obtained by DMRG.\cite{9}  Thus while there is general 
agreement that this ground state is a spin liquid, the precise nature of the spin liquid 
remains somewhat unsettled.

Experimentally it is known that about 15\% of the Zn ($S=0$) ions which are located 
between the Kagome planes are replaced by $S = 1/2$ Cu ions.  It has been argued 
that there is not much Zn substitution for Cu in the Kagome planes,\cite{10} so that 
the disturbance of the Kagome structure may be minimal.  However, much of the low 
energy excitations measured by thermodynamic probes such as specific heat and spin
susceptibility are dominated by the local moments between planes.  Furthermore, due 
to spin orbit coupling, we expect deviation from the Heisenberg model.  To first order 
in the spin orbit coupling constant $\lambda$, we expect Dzyaloshinskii-Moriya (DM) 
terms of the form
\begin{equation}
{\cal H}_{\rm DM} = 
\sum_{\langle ij \rangle} \bm {D}_{ij} \cdot \bm {S}_i \times \bm {S}_j
\end{equation}
where the DM vector $\bm{D}_{ij}$ is located on bond $\langle ij \rangle$.  
Since $\bm {D}_{ij} = -\bm {D}_{ji}$, the $\bm {D}_{ij}$ vectors depend on the convention of 
the bond orientation.\cite{11}  For a given convention the DM vectors are shown in Fig.~1.  
There is also an exchange anisotropy term $J_{zz} \neq J_{xx}$.  However, the difference 
is second order in $\lambda$ and may be less important than the DM term.  
The out-of-plane DM term $(D^z)$ has been estimated to be about 8\% of $J$. 
Due to the delicate nature of the ground state of the Heisenberg model explained above, 
it is not at all clear that the nearest-neighbor Heisenberg result applies to the Herbertsmithite.
\begin{figure}[t]
\centerline{
\includegraphics [width=3.2in]{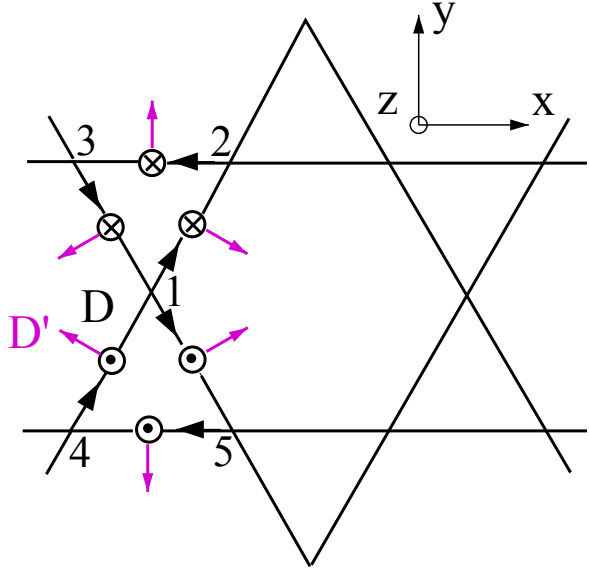}
}
\caption{The DM vectors on the Kagome lattice for Herbertsmithite.  
The arrow specifies the order of the operator $\bm{S}_i \times \bm{S_j}$. 
(adapted from Ref. 11)  $D$ is the $z$ component while $D^\prime$ is the in-plane component.}

\end{figure}

The defining character of a quantum spin liquid is the emergence of exotic particles such 
as spinons which carry $S = 1/2$ and the associated gauge fields.\cite{12}  In the U(1) spin liquid, 
the gauge field is gapless whereas in the Z$_2$ spin liquid the gauge field is gapped.  
The gauge field is defined by the phase $a_{ij}$ of the spinon hopping matrix element $te^{ia_{ij}}$ on link $ij$.
It is a compact gauge field and the spin liquid corresponds to the deconfined phase of the 
gauge field, so that the compactness may be ignored in the long wavelength limit and $a_{ij}$ 
may be replaced by a continuum field $\bm{a}(\bm{r})$.  The gauge invariant quantities are 
the gauge field $\bm{b}= \bm{\nabla} \times \bm{a}$ which is in the $\hat{z}$ direction and 
the gauge electric field $\bm{e} = -\bm{\nabla}a_0 + {d\bm{a}\over dt}$ which lies in the plane.  
The physical meaning of the magnetic flux has been extensively discussed.\cite{13}  If $\Phi$ 
is the flux through a plaquette, 
$\sin\Phi$ is one half of the solid angle subtended by the spins along the plaquette.  
For a three site triangle, we have
\begin{equation}
\sin \Phi = {1\over 2} \bf{S}_1 \cdot (\bm{S}_2 \times \bm{S}_3)
\end{equation}
The quantity $\bm{S}_1 \cdot \bm{S}_2 \times \bm{S}_3$ is known as the scalar spin chirality.  
Thus the fluctuation spectrum of the gauge magnetic field is proportional to the fluctuation of 
the spin chirality.  It is highly desirable to measure this correlation function because it gives 
information on the gauge fluctuation and can help to distinguish different spin liquids.  
However the  measurement of a correlation of a product of 3 spin operators is a daunting task.  
A method to measure the vector chirality $\bm{S}_1 \times \bm{S}_2$ has been suggested by Maleev (ref 14), but that does not apply to scalar chirality. Shastry and Shraiman (ref.15) have suggested measuring chirality fluctuations using Raman scattering, but that contains information only for very small {\em q}. 
A proposal to measure this using resonant X-ray scattering has been made.\cite{16}  However, 
the energy resolution of this technique is currently limited to 20~meV or so, which is on the scale of $J$.  

It turns out that in the Kagome lattice we can turn the DM term to our advantage and achieve a simpler 
measurement of the chirality correlation.  The hint comes from a recent paper by Savary and Balents,\cite{17} 
who pointed out that in a certain pyrochore spin-ice material, the gauge  field (in their case the electric 
field\cite{18}) is proportional to $S_z$ and its fluctuation can be directly measured by neutron scattering.  
This system is treated in the strong spin-orbit coupling where $J$ rather than spin $S$ is a good quantum 
number, but this work raises the possibility of finding the same proportionality in the presence of weak 
spin-orbit coupling.  We find that it is indeed the case for the Kagome lattice with the DM 
vectors shown in Fig.~1.  In the Kagome lattice each site is connected to two triangles.  
Let us consider the site labeled 1. The total chirality through the two attached triangles is
\begin{equation}
\chi_1 = \bm{S}_1 \cdot  \bm{S}_2 \times \bm{S}_3 + \bm{S}_1 \cdot \bm{S}_4 \times \bm{S}_5
\end{equation}
However, due to the DM term,
\begin{eqnarray}
\langle  \bm{S}_2 \times \bm{S}_3  \rangle & = & \alpha \bm{D}_{23}  \nonumber \\
\langle  \bm{S}_4 \times \bm{S}_5  \rangle & = & \alpha \bm{D}_{45}  =  
-\alpha \bm{D}_{54}  =  \alpha \bm{D}_{23}
\end{eqnarray}
where $\alpha$ is a constant and the last step is by inspection from Fig.~1.  
For $|\bm{D}_{23}| \ll J$, we expect $\alpha \sim {1\over J}$.  The important point is 
the contributions from the two triangles add and we find a linear coupling between 
$\chi_1$ and $\bm{S}_1 \cdot \bm{D}_{23}$.  If we average over corners of the triangle 
$123$ and define $\bar{\chi} = {1\over 3}(\chi_1 + \chi_2 + \chi_3)$, it is clear that the in-plane component 
of the $\bm{D}$ vectors cancel  and a local fluctuation in $S_z({\bm r},t)$ induces a local 
fluctuation in $\bar{\chi}(\bm{r},t)$ with a proportionality constant of $(2 \alpha D^z_{23})$.  
This suggests that a measurement of the $\langle S_z (r,t)S_z(0,0) \rangle$ correlation function 
will contain a piece which is proportional to the chirality correlation $\langle \bar{\chi}(r,t)\bar{\chi}(0) \rangle$.  
A more formal argument proceeds as follows.  Let $|\alpha\rangle$ and $E(\alpha)$ denote the exact 
eigenstates and energies of the Hamiltonian $H_0$ without the DM term and we perturb in $H_{\rm DM}$.  
We are interested in a subset of the excited states, denoted by $|\alpha_\chi \rangle$ which are connected 
to the ground state $|0\rangle$ by the operator $\chi_1$, i.e., $\langle \alpha_\chi |\chi_1| 0 \rangle \neq 0$.  
On the other hand, the operator $S_z(\bm{r})$ has no matrix element to these states, i.e., 
$\langle \alpha_\chi |S_z(\bm{r})| 0 \rangle = 0$.  The gauge magnetic field spectral function is 
proportional to the chirality spectral function
\begin{equation}
S_\chi (q,\omega) = \sum_{\alpha_\chi}
\left|
\langle \alpha_\chi | \chi_{\bm{q}} | 0 \rangle
\right|^2
\delta 
\left
(\omega - (E(\alpha_\chi) - E(0)\right)
\end{equation}
Now we turn on the Dirac terms.  The neutron scattering cross-section is proportional to
\begin{equation}
S(q,\omega) = \sum_f
\left|
\langle f | S_z(\bm{q}) |i \rangle
\right|^2
\delta (\omega - (E_f - E_i))
\end{equation}
where $| f \rangle$, $E(f)$ are the exact eigenstate and energy of the total Hamiltonian.  
We now compute the correction to the matrix element $\langle f | S_z |i\rangle$ to first order 
in $H_{\rm{DM}}$ for the state $| f_\chi \rangle$ which derives from the state $| \alpha_\chi\rangle$, i.e.,
\begin{equation}
| f_\chi \rangle = | \alpha_\chi \rangle +
\sum_{\alpha \neq \alpha_\chi}
{\langle \alpha | H_{\rm{DM}} | \alpha_\chi \rangle \over E(\alpha_\chi) - E(\alpha)}
|\alpha \rangle
\end{equation}
and similarly for $| i \rangle$.  Since the zeroth order term vanishes, we find
\begin{equation}
\langle f_\chi | S_z(\bm{r}_1) | i \rangle =
\sum_\alpha 
\left[ {\langle \alpha_\chi | H_{\rm{DM}} |\alpha \rangle \langle\alpha |S_z|0\rangle \over E(\alpha_\chi) - E(\alpha)}
+ {\langle\alpha _\chi |S_z|\alpha \rangle \langle \alpha |H_{\rm{DM}} |0 \rangle
\over E(0) - E(\alpha)} \right]
\end{equation}
Since 
$| 0 \rangle$ and $| \alpha_\chi \rangle$ 
are total spin singlets, we argue that by choosing a different spin quantization axis, $S_z$ can rotated to $S_x$ 
in Eq.(8) and $|\alpha \rangle$ are spin triplet states.  If the spin triplets have a large gap $\Delta_t$ we may replace 
$E(0) - E(\alpha)$ in the second term by $-\Delta_t$.  If we are interested in low energy modes of the chirality 
fluctuations such that $\omega = E(\alpha_\chi) - E(0) \ll E(\alpha) - E(0) \approx \Delta_t$, we may likewise 
replace that energy denominator in the first term by $-\Delta_t$.  Then the sum over $|\alpha \rangle$ can be 
done and it is clear that
\begin{equation}
\langle f_\chi | S_z(\bm{r}_1) |i\rangle = 
-\sum_{\langle jk \rangle} {2 D_{jk} \over \Delta_t}
\langle \alpha_\chi | S_z (\bm{r}_1) \hat{z} \cdot \bm{S}(\bm{r}_j) \times \bm{S}(\bm{r}_k)|0\rangle
\end{equation}
where the sum is over sites $j,k$ which are connected to site 1 as corners of a triangle, i.e.,
$(j,k)=(2,3)$ or $(4,5)$.
Putting this matrix element into Eq.(6) we see that neutron scattering indeed has a 
piece which couples to chirality fluctuations, with a form factor given by
$
|
\sum_{\langle jk \rangle}
{2D_{jk} \over \Delta_t}
|^2
$,
at least in the limit where the triplet gap is large.  We expect this coupling will continue to hold 
even if the assumption of a large $\Delta_t$ fails, as in the Dirac spin liquid case, but the form 
factor may acquire some quantitative differences and perhaps some $\bm{q}$ and $\omega$ 
dependence. This will be the case if the contribution to the sum in Eq(8) from low energy (near gapless) 
triplet excitations are reduced due to restricted phase space and we can approximately replace the 
energy denominator by an average triplet excitation energy.

As an example, if the ground state is described by the U(1) Dirac spin liquid, the correlation of the gauge 
magnetic field is expected to be (in the RPA approximation\cite{19})
\begin{equation}
\langle
| b(q,\omega)|^2 
\rangle
\sim
{q^2 \theta (\omega-vq) \over (\omega^2-v^2q^2)^{1/2}}
\end{equation}
where $\theta$ is the step function and $v$ is spinon velocity.  The neutron scattering intensity 
which couples to the $S_z$ channel is expected to have a piece given by Eq.(10), with the intensity 
reduced by $(2\alpha D^z_{ij})^2$.  If the ground state is a Z$_2$ gapped spin liquid, the gauge fluctuation 
is gapped and one may expect to see a gapped mode instead of Eq.(10).  As pointed out by Savary and 
Balents,\cite{17} the $q^2$ dependence in Eq.(10) implies that the gauge fluctuation structure factor 
vanishes as $q \rightarrow 0$, in contrast with the spin wave structure factor which diverges as $1/q$ 
in this limit.  While this helps identify the gauge fluctuation, it also makes its detection more difficult.

\begin{figure}[h]
\centerline{
\includegraphics [width=3.2in]{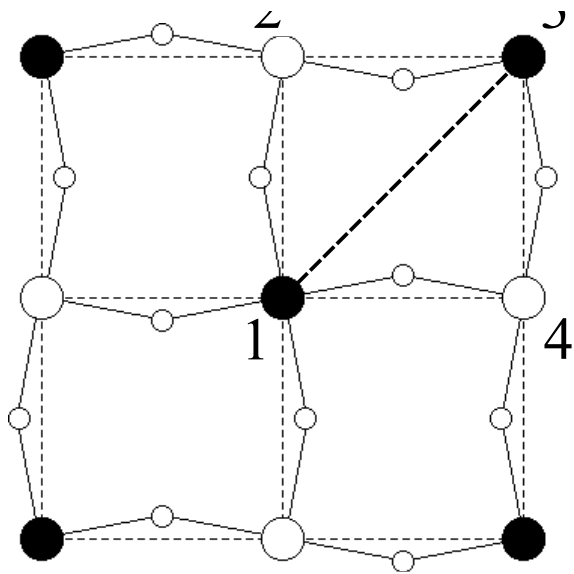}
}
\caption{The distortion of the Ir (solid) and oxygen (open) ions in the Ir-0 plane of Sr$_2$IrO$_4$. 
(adapted from Ref. 20)}
\end{figure}

The linear relationship between chirality fluctuation and $S_z$  fluctuation is rather special to the 
Kagome lattice.  Such a coupling does not exist for the square lattice, for instance, for slowly varying chirality fluctuations. Consider magnetic 
ions on a square lattice each surrounded by corner sharing oxygen cages which rotate about the $\hat{z}$ 
axis in a staggered manner. (Sr$_2$IrO$_4$ is such an example\cite{20} shown in Fig.2.) 
In this case each magnetic ion is connected to 8 triangles whose opposite side is a nearest-neighbor bond 
with a DM vector in the $\hat{z}$ direction.  The gauge magnetic flux through each square is given in terms 
of the scalar chirality of the triangles which split up the square.\cite{13}  It is easy to see that the contributions 
from the two triangles 123 and 134 which split up a square cancel each other and there is no linear coupling 
between the gauge flux and $S_z$ in this case.

In conclusion, we have shown that in the Kagome lattice, the DM term leads to a linear coupling 
between the $S_z$ fluctuation with the spin chirality fluctuation. It will be interesting to see if neutron 
scattering can give information on the chirality fluctuation and shed light on the nature of the spin liquid in 
Herbertsmithite.

We thank Leon Balents for helpful discussions. We also thank the KITP for its hospitality. PAL acknowledges
support by the NSF through DMR 1104498.
NN is supported by Grant-in-Aid
for Scientific Research (Grants No. 24224009) from
the Ministry of Education, Culture, Sports, Science and
Technology of Japan, Strategic International Cooperative
Program (Joint Research Type) from Japan Science
and Technology Agency, and Funding Program for
World-Leading Innovative RD on Science and Technology
(FIRST Program)


\begin{references}
\bibitem{18}
Due to duality the naming of electric and magnetic fields is a matter of choice.  In the choice made in Ref. 15, the electric field is odd under time reversal and has the symmetry of the more conventional choice for magnetic field.
\end{references}
\end{document}